\documentclass[aps,prb,twocolumn,showpacs,superscriptaddress]{revtex4}
\usepackage{hyperref}
\usepackage{natbib}
\usepackage{graphicx}

\begin{document}

\title{Magnetic field of Josephson vortices outside superconductors}
\author{V. G. Kogan}
\affiliation{ Ames Laboratory and Department of Physics and Astronomy, ISU,
Ames, IA 50011 }
 \author{V. V. Dobrovitski }
 \affiliation{ Ames Laboratory and Department of Physics and Astronomy,
 ISU, Ames, IA 50011 }

\date{\today}
\begin{abstract}
We consider the structure of Josephson vortices approaching the junction
boundary with vacuum in large area Josephson junctions with  the
Josephson length $\lambda_J$ large relative to the London penetration
depth $\lambda_L$. Using the stability argument for 
one-dimentional solitons with respect to 2D perturbations, it is shown
that on the scale 
$\lambda_J$ the Josephson vortices do not spread near the boundary in the
direction of the junction.
The field distribution in vacuum due to the Josephson vortex is evaluated, the
information needed for the Scanning SQUID Microscopy.\end{abstract}


\maketitle

A new experimental technique, the scanning SQUID microscopy (SSM), has
recently been developed for measuring magnetic field distributions due to 
 vortices exiting superconducting samples.\cite{Kirtley} Knowing the
distributions one can, in principle, extract the London penetration depth
$\lambda_L$ (either isotropic or anisotropic) and its temperature
dependence.\cite{KirtleyT} 

To extract the superconducting parameters from the measured field above the
sample one has to solve the problem of the field created by a vortex
terminating at the sample surface inside and outside the sample. For
isotropic bulk materials the solution has been given by 
J.~Pearl who utilized the cylindrical symmetry of a vortex normal to the sample surface.\cite{Pearl} For the
anisotropic case, a more general approach should be employed to match
solutions of London equations inside and of Maxwell equations
outside.\cite{KSL} The most likely situation of a vortex 
perpendicular to the sample surface  was described and
implemented in analysis of SSM data in Ref.\,\onlinecite{KKCM}. These
calculations showed that  vortices spread  out in the superconductor while 
approaching the surface.  For the vortex axis along $z$, the transverse
components $h_{x,y}$ appear describing  field lines bending out of the
vortex axis in the subsurface layer of a depth on the order of 
$\lambda_L$.  

Originally, the SSM technique has been designed for  
 studies of Josephson boundaries between misoriented  high-$T_c$ superconducting
crystallites. Some of the most convincing demonstrations of the $d$-wave
symmetry of these materials were obtained by using SSM to detect  
spontaneous half-flux-quantum vortices at the intersection of three Josephson boundaries between  crystallines of a proper
misorientation.\cite{Kirtley2,miller} Other configurations involving
Josephson boundaries have been shown to contain  spontaneous magnetic
flux (e.g., closed triangular and hexagonal boundaries,\cite{Kirtley4}  faceted boundaries with alternating critical current\cite{***}). However, absence of theoretical description of Josephson vortices terminating on the sample surface prevented a reliable interpretation of a large body of SSM data. 

The Josephson length $\lambda_J$ for the boundaries in question (the scale of the field distribution in the Josephson vortex  
that plays a role of $\lambda_L$ for the bulk vortices) might be on
the order of microns. This suggests that a considerable
spreading effect may take place at the distance of order of $\lambda_J$
near the surface. Such an effect, if exists, should be taken into account when relating the outside field distributions to the internal structure of Josephson vortices (J-vortices). 

We argue in this communication that Josephson vortices do not spread in the direction of the junction in the limit 
\begin{equation}
\lambda_L\ll\lambda_J\,. \label{L<<J}
\end{equation}
 This conclusion makes the SSM data
analysis not only possible but quite straightforward.  

Let us start with an infinite two-dimensional (2D) Josephson junction in the plane $(x,z)$ between
two superconducting banks occupying the half-spaces $y>0$ and $y<0$. The
stationary J-vortex directed along $z$ in such a junction is
described by the sine-Gordon equation  
\begin{equation}
\lambda_J^2\,\varphi^{\prime\prime}=\sin\varphi \,, 
\label{sinG}
\end{equation}
where $\varphi (x)$ is the gauge-invariant phase
difference.\cite{Kulik,Barone,Tinkham}   The Josephson length
$\lambda_J\propto  j_c^{-1/2}$ may vary due to differences in crystals
misalignment and in quality of the boundary ($j_c$ is the critical
Josephson current density). The primes denote differentiation with respect to $x$.  One follows the standard procedure: multiply (\ref{sinG}) by
$\varphi^{\prime}$ to get the first integral
$\lambda_J^2\varphi^{\prime 2}/2=C-\cos\varphi$. The Josephson current
and the field ($\propto \varphi^{\prime}$) at the junction should vanish
as $|x|\rightarrow \infty$; this yields $C=1$ for $\varphi(\infty)=2\pi$.
We then obtain:
 \begin{equation}
\varphi = 4\tan^{-1}(e^{ x/\lambda_J})\,.
\label{soliton}
\end{equation}
The field at the junction plane: 
\begin{equation}
h_z( x)={\phi_0\over 4\pi \lambda_L}\,{d\varphi\over dx }=\frac{\phi_0}{ 
2\pi \lambda_L\lambda_J}\, {\rm sech} {x\over\lambda_J}\,;
\label{field}
\end{equation}
 $\phi_0=\pi\hbar c/|e|$ is the flux quantum  
(the thickness of the insulating layer is assumed small relative to
$\lambda_L$). The vortex described by Eqs. (\ref{soliton}) and
(\ref{field}) is infinite in the $z$ direction; the problem and the solution are, in fact, one-dimensional.

Let us now remove the superconductors from the half-space $z>0$; thus
  the plane $z=0$ is a free boundary  with vacuum. The question arises 
 what changes the J-vortex (occupying now only $z<0$) should 
undergo? The problem is no longer uniform in the $z$ direction, and the $z$  independence of the solution of Eq.\,(\ref{sinG}) cannot be assumed. In other words, we have now $\varphi=\varphi(x,z)$ and the sine-Gordon equation  in two spacial dimensions: 
\begin{equation}
\lambda_J^2\,\nabla^2\varphi =\sin\varphi \,, 
\label{sinG2}
\end{equation}
 $\nabla^2$ is the 2D Laplacian. We look for solutions at the half-plane $-\infty<z<0$ describing a J-vortex  somehow spreading as $z\to -0$. 
The solution  should satisfy certain boundary conditions:
at $z\rightarrow -\infty$, $\varphi$ must be of the form
(\ref{soliton}). Besides, at any $z<0$, the vortex field should vanish far from the vortex ``core": 
\begin{equation}
 \frac{\partial\varphi}{\partial x}\Big|_{x\rightarrow \pm\infty}=0\,.
\label{df/dx=0}
\end{equation}

 This formidable, at first sight, problem does have an exact and unique solution  due to the ``nonexistence" theorem by Derrick \cite{Derrick,Eilenberger} that 
states that there are no {\it stable static} solutions of 
the sine-Gordon equation in more than one spatial dimension. 

We  reproduce this argument for the 2D case of interest here.
The energy functional  $E\{\varphi (x,z)\}$, which generates  Eq.\,(\ref{sinG2}),  can be written in the form:
\begin{equation}
 E=\int \int dx\,dz\,\Big[(\nabla\varphi )^2+\frac{4}{\lambda_J^2}\sin^2\frac{\varphi}{2}  \Big]=K+U\,,
\label{functional}
\end{equation}
where we have dropped a constant  prefactor irrelevant for the following. 
Let us assume that a function $\varphi (x,z)$ exists such that $\delta E/\delta\varphi=0$; in other words, that the sine-Gordon equation (\ref{sinG2}) has a 2D bounded solution. We now evaluate the energy for the phase $\varphi (\eta x,\eta z)$ with an arbitrary scaling factor $\eta$. Substitute this in the integral (\ref{functional}) and change integration variables to $\eta x,\eta z$ to obtain:
\begin{equation}
 E(\eta)= K+U/\eta^2\,,
\label{eq8}
\end{equation}
where the kinetic and potential parts, $K$ and $U$, are defined for the original phase $\varphi (x,z)$. 
Since $\varphi (\eta x,\eta z)$ is a solution of $\delta E=0$ for $\eta=1$, we must have $dE(\eta)/d\eta |_{\eta=1}=0 $. Equation (\ref{eq8}) however gives 
$dE(\eta)/d\eta |_{\eta=1}=-2U<0$.  Thus, there are no 2D static solutions--either stable or unstable--of Eq.\,(\ref{sinG2}). 

Note that for 1D case $E(\eta)= \eta K+U/\eta$, $dE(\eta)/d\eta |_{\eta=1}=K-U$ and 1D solutions with $K=U$ do exist and are stable.

It should be noted that the above argument implies an infinite plane $x,z$;  both integrals (\ref{functional}) are taken on the interval $(-\infty,\infty)$. For the junction of our interest $-\infty<z<0$. Clearly, the scaling employed above can still be used for the half-plane provided the coordinate $z$ is counted from the junction edge so that the edge is left in place under the scaling transform. 

The variational minimization of the energy (\ref{functional}) defined on the half-plane involves the integration by parts which yields the boundary condition  
\begin{equation}
 \frac{\partial\varphi}{\partial z}\Big|_{z\rightarrow  -0}=0\,,
\label{df/dz=0}
\end{equation}
which is invariant with repect to the scaling employed above. This condition translates to $h_x(x,0,0)=0$ along with $g_z(x,+0,-0)=g_y(x,+0,-0)=0$.
 
We thus conclude that Eq.\,(\ref{soliton}) with $z$ independent $\varphi(x)$ is the only solution possible. In other words, the static 
Josephson vortex does not spread approaching the sample surface. 

It is worth mentioning that dynamic 2D solutions of the time dependent sine-Gordon
equation do exist. A few examples are given in Ref.\,\onlinecite{christ}
where the initial essentially two-dimensional soliton-like structure
 has been shown (numerically) to evolve in time toward a
straight 1D soliton. 

We stress that the above conclusion holds when effects on the scale of
London penetration depth are neglected. Of course, at distances of the
order $\lambda_L$ from the surface, the field deviates from the $z$
direction and some spreading takes place. This effect, however, is
hardly relevant for the SSM method because the low bound of the spatial
resolution of the SQUID loops  is set by the loop size, which
exceeds by much the London depth.\cite{Kirtley}

Also, one has to keep in mind  that near the free surface, the stray fields 
outside the superconductor may affect the very equation governing the behavior 
of the phase difference $\varphi$. This question has been discussed for 
thin-film junctions,\cite{Mints_1,Ivanchenko,JJF}  where it was shown that 
the governing equation replacing the ``bulk" Eq.\,(\ref{sinG2}) becomes 
integral, in other words, the stray fields make the problem nonlocal. 
This perturbation, however, penetrates the system only down to depths on 
the order of $\lambda_L$ and can be disregarded in our approximation.

\begin{figure}[htb]
\begin{center}
\includegraphics[width=6.cm ]{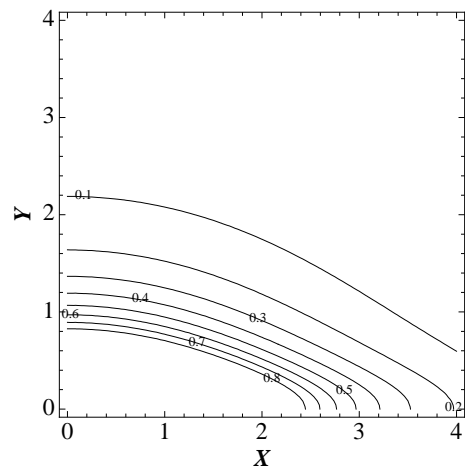} 
\caption{Contours of $h_z(x,y)=$ const for $\lambda_J=1$ and  $z_0=0.5 $. The horizontal axis is $x$ along the junction. The field in units of $\phi_0/4\pi^2 \lambda_J^2$ reaches maximum at the origin $h_z(0,0)=3.507 $. Values of $h_z$ are indicated on the contours.}
\label{fig1} 
\end{center}
\end{figure}

\begin{figure}[htb]
\begin{center}
\includegraphics[width=6.cm]{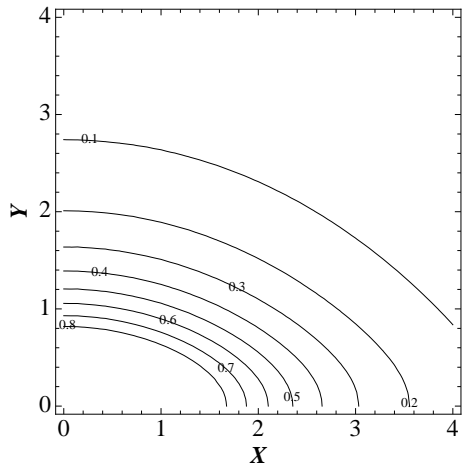}
\caption{Same as Fig.\,\ref{fig1} for $z_0=1$.  The maximum field is   $h_z(0,0)= 1.474$.}
\label{fig2} 
\end{center}
\end{figure}

\begin{figure}[htb]
\begin{center}
\includegraphics[width=6.cm]{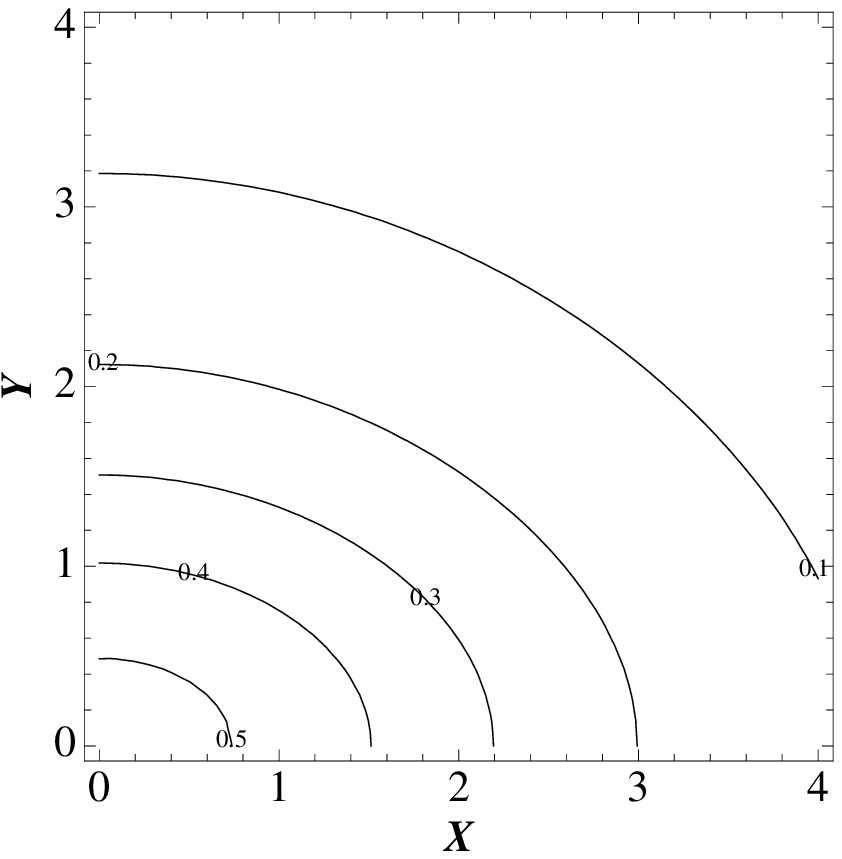} 
\caption{Same as Fig.\,\ref{fig1} for $z_0=2$.  The maximum field is   $h_z(0,0)= 0.5388$.}
\label{fig3}
\end{center}
\end{figure}

Since the field ${\bf h}$ in vacuum satisfies  
${\rm curl}\,{\bf h}={\rm div}\,{\bf h}=0$, one can introduce in the upper half-space $z>0$ a scalar potential:
\begin{equation}
{\bf h}=\nabla \psi\,,\qquad \nabla^2\psi =0\,.
\label{Laplace}
\end{equation}  
The boundary conditions for the Laplace equation are provided by the
Josephson field ``sticking out" of the plane $z=0$ from a belt of the width $2\lambda_L$ along the junction. With the help of Eq.\,(\ref{field})
we obtain:
\begin{equation}
 \frac{\partial\psi(x,y,z)}{\partial z}\Big|_{z=0} =
2\lambda_L\,\delta(y)\,\frac{\phi_0}{2\pi  
 \lambda_L\lambda_J}\, {\rm sech} {x\over\lambda_J}\,.
\label{source}
\end{equation}
In magnetostatic terms, this corresponds to a linear ``charge" with the 
density 
\begin{equation}
-\frac{\phi_0}{4\pi^2\lambda_J}\,  {\rm sech} {x\over\lambda_J}\,.
\label{charge}
\end{equation}
Therefore, the potential $\psi$ reads:
\begin{equation}
\psi(x_0,y_0,z_0) =-\frac{\phi_0}{4\pi^2 \lambda_J}
\int_{-\infty}^{\infty}\frac{dx}{R\,\cosh(x/\lambda_J)}\,, 
\label{potential}  
\end{equation}
where
\begin{equation}
R^2= (x-x_0)^2+y_0^2+z_0^2 \,.
\label{R}
\end{equation}
 The field
component $h_z$, which is actually measured by SSM, is readily obtained:
\begin{equation}
h_z(x_0,y_0,z_0) =\frac{\phi_0\,z_0}{4\pi^2 \lambda_J}
\int_{-\infty}^{\infty}\frac{dx}{R^3\,\cosh
(x/\lambda_J)}\,.
\label{result}
\end{equation}
This  integral is well convergent at any finite $z$ and is easy for numerical evaluation. It diverges at the junction line at $z=0$; the divergence, however, should be truncated at distances of the order $\lambda_L$ which are disregarded in our model.   
  
Contours of $h_z(x,y)=const$ are shown in Figs.\,\ref{fig1}-\ref{fig3}.  For this calculation we set $\lambda_J=1$ and  
   $z_0=0.5,1$ and 2.  

In actual SSM, the field $h_z(x,y;z)$ at a fixed height $z$
should be integrated over the SQUID area. In some implementations  it is more
convenient to have the 2D Fourier transform $h_z(k_x,k_y;z)$ instead of
the real space distribution. After straightforward algebra we obtain:
\begin{equation}
  h_z(k_x,k_y;z) =\frac{\phi_0\, e^{-kz}} { \cosh
(\pi k_x\lambda_J/2)}\,,\quad k=\sqrt{k_x^2+ k_y^2}\,.
\label{FT}
\end{equation}
In particular this shows  that the total flux through any plane $z=z_0$
is $h_z({\bf k}=0;z_0)=\phi_0$.\\

We thank J. R. Kirtley and J. R. Clem for helpful discussions. 
Work at the Ames Laboratory is supported by the Department of
Energy Office of Basic Energy Sciences under Contract No. DE-AC02-07CH11358.


\begin{thebibliography}{99}

\bibitem{Kirtley} J.R. Kirtley,  C.C. Tsuei, M. Rupp,  J.Z. Sun, Lock See
Yu-Jahnes, A. Gupta, M.B. Ketchen, K.A. Moler, and M. Bhushan,
Phys. Rev. Lett. {\bf 76}, 1336 (1996).
 
\bibitem{KirtleyT} J.R. Kirtley, C.C. Tsuei,  K.A. Moler, V.G. Kogan, J.R.
Clem, A.J. Turberfield,
Appl. Phys. Lett. {\bf 74}, 4011 (1999).

\bibitem{Pearl}J. Pearl, J. Appl. Phys. {\bf 37},4139 (1966)

\bibitem{KSL}V. G. Kogan, A. Yu. Simonov, and M. Ledvij,
Phys. Rev. B {\bf 48}, 392 (1993).

\bibitem{KKCM} J. R. Kirtley,  V. G. Kogan, J. R. Clem, and K. A. Moler,
 Phys. Rev. B, {\bf 59}, 4343   (1999).

\bibitem{Kirtley2} C.C. Tsuei, J.R. Kirtley, C.C. Chi, L.S. Yu-Janes, A. Gupta, T. Shaw, J.Z. Sun, and M.B. Ketchen, 
Phys. Rev. Lett. {\bf 73}, 593 (1994).

\bibitem{miller} J.H. Miller, Jr., Q.Y. Ying, Z.G. Zou, N.Q. Fan, J.H. Xu,
M.F. Davis, and J.C. Wolfe, 
Phys. Rev. Lett. {\bf 74}, 2347 (1995).

\bibitem{Kirtley4}J.R. Kirtley, P. Chaudhari, M.B. Ketchen, N. Khare,
Shawn-Yu Lin, and T. Shaw, Phys. Rev. B, {\bf 51}, 12057 (1995).

\bibitem{***}R. G. Mints and  V. G. Kogan,   Phys. Rev. B, {\bf 55}, R8682 (1997).

\bibitem{Kulik}I.O. Kulik and I.K. Yanson, {\it The Josephson Effect in
Superconductive Tunneling Structures}, Jerusalem, Israel Program for Scientific Translations, 1972.

\bibitem{Barone} A. Barone and G. Paterno, {\it Physics and Applications of the
Josephson Effect}, John Willey, New York, 1982.

 \bibitem{Tinkham}M. Tinkham, {\it Introduction to Superconductivity},
New York, McGrow-Hill, 1996.

\bibitem{Derrick}G.H. Derrick, J. Math. Phys. {\bf 5}, 1252 (1964).

\bibitem{Eilenberger}  G. Eilenberger, {\it Solitons}, Springer Series in Solid State
Sciences, v.19, Springer-Verlag, Berlin, 1981.

\bibitem{christ} P.L. Christiansen and P.S. Lomdahl, Physica {\bf 2D}, 482 
   (1981).

\bibitem{Mints_1} R. G. Mints and I. B. Snapiro, \prb{\bf 49}, 6188 (1994);
  \prb{\bf 51}, 3054 (1995);
 \prb{\bf 52}, 9691 (1995).
 
\bibitem{Ivanchenko} Y. M. Ivanchenko, \prb{\bf 52}, 79 (1995).

\bibitem{JJF} V. G. Kogan, V. V. Dobrovitski, J. R. Clem, Y. Mawatari, and R. G.
Mints, Phys. Rev. B, {\bf 63}, 144501   (2001).


\end{thebibliography}
\end{document}